\documentclass[aps,prd,draft,showpacs]{revtex4}

\begin{document}

\title{Effective Action for the Scalar Field Theory with Higher Vertices}
\author{Chungku Kim}
\date{\today}

\begin{abstract}
We derive a new kind of recursion relation to obtain the
one-particle-irreducible (1PI) Feynman diagrams for the effective action. By
using this method, we have obtained the graphical representation of the
four-loop effective action in case of the general bosonic field theory which
have vertices higher than the four-point vertex.
\end{abstract}
\maketitle


\affiliation{Department of Physics, College of Natural Science, Keimyung
University, Daegu 705-701, KOREA}

\section{Introduction}

In quantum field theory, the effective action plays an important role in
studies of the vacuum instability, the dynamical symmetry breaking and the
dynamics of composite particles\cite{Sher}. It is well known that the
effective action of the given particle physics model can be obtained from
the 1PI vacuum diagrams with the generalized propagator and the vertices
which depend on the classical field \cite{Jackiw}. There exist various
packages such as FeynArts\cite{F} and QGRAF\cite{Q} to determine the Green
functions of the given particle physics model. Recently, a systematic
approach to obtain the recursive generation of the connected and the 1PI
Feynman diagrams of the multicomponent $\phi ^{4}$-theory, QED and the
scalar QED was proposed by using the functional integral identity $\int
D\Phi \ \frac{\delta }{\delta \Phi }F[\Phi ]=0$ \cite{R2}\cite{R3} \cite{R4}%
\cite{R5}\cite{R6}\cite{R7}\cite{R8} . Moreover the recursive generation of
the two-particle-irreducible (2PI) effective action have been analyzed\cite
{2PI} and the four-particle- irreducible (4PI) effective action was obtained
by using the result of the 2PI effective action\cite{4PI}. In this paper, we
propose a new kind of recursion relation to obtain the 1PI Feynman diagrams
for the effective action. In Sec.II, we derive the recursion relation for
the effective action and apply this method to the general bosonic field
theory which have vertices higher than the four-point vertex and obtain the
graphical representation of the four-loop effective action. In Sec.III, we
give some discussions and conclusions.

\section{A\ New Recursion Relation for the Feynman Diagrams of the Effective
Action}

In this section, we will first derive a recursion relation for the Feynman
diagrams of the effective action for the action given by

\begin{equation}
S[\Phi ]=\int \{\frac{1}{2}\Phi _{A}\Delta _{AB}^{-1}\Phi _{B}+S^{int}[\Phi
]\}.
\end{equation}
where the interaction $S^{int}[\Phi ]$ contains the higher vertices which
appear in lattice regularization\cite{lattice} as well as the cubic and the
quartic interactions. In this paper, we use a notation where the capital
letters contain both the space-time variables and the internal indices and
the repeated capital letters mean both the integration over continuous
variables and the sum over internal indices. For example, if the capital
letter $A$ contains a space-time variable $x$ and the internal index $i,$

\begin{equation}
J_{A}\Phi _{A}\equiv \sum_{i}\int d^{4}xJ_{i}(x)\Phi _{i}(x).
\end{equation}
The generating functional for the Green functional $W[J]$ is given by the
functional integral

\begin{equation}
\exp \{-\frac{1}{\hbar }W[J]\}=\int D\Phi \ \text{exp}\{-\frac{1}{\hbar }%
(S(\Phi )-J_{A}\Phi _{A})\}.
\end{equation}
Here $\hbar $ is an expansion parameter and we will put $\hbar =1$ at final
stage. The effective action $\Gamma [\phi ]$ is defined by the Legendre
transformation of the Green functional $W[J]$ as 
\begin{equation}
\Gamma [\phi ]=W[J]-J_{A}\phi _{A},
\end{equation}
where

\begin{equation}
\phi _{A}\equiv \frac{\delta W[J]}{\delta J_{A}}.
\end{equation}
By using (4) and (5), we can obtain the relation 
\begin{equation}
\frac{\delta \Gamma [\phi ]}{\delta \phi _{A}}=-J_{A}.
\end{equation}
and from (3) and (4), we can write

\begin{equation}
\exp \{-\frac{1}{\hbar }\Gamma [\phi ]\}=\int D\Phi \ \text{exp}\{-\frac{1}{%
\hbar }(S(\Phi )-J_{A}(\Phi _{A}-\phi _{A}))\}.
\end{equation}
By expanding the effective action $\Gamma [\phi ]$ around $\hbar $ as

\begin{equation}
\Gamma =\sum_{l=0}\hbar ^{l}\Gamma ^{(l)}[\phi ],
\end{equation}
, we can obtain the loop-wise expansion of $\Gamma [\phi ]$\cite{Coleman}.
Now let us change the variable of the functional integral $\Phi \rightarrow $
$\Phi +\phi $ and expand $S(\Phi +\phi )\ $as

\begin{equation}
S[\Phi +\phi ]=S[\phi ]+\sum_{N=1}\frac{1}{N!}S_{A_{1\cdot \cdot \cdot
}A_{N}}[\phi ]\Phi _{A_{1}}...\Phi _{A_{N}},
\end{equation}
where 
\begin{equation}
S_{A_{1\cdot \cdot \cdot }A_{N}}[\phi ]\equiv \frac{\delta ^{N}S[\phi ]}{%
\delta \phi _{A_{1}}\cdot \cdot \cdot \delta \phi _{A_{N}}}.
\end{equation}
Actually the vertex $S_{A_{1\cdot \cdot \cdot }A_{N}}$ corresponds to one
point in space-time. By substituting (9) into (7), we can obtain the first
two terms of the effective action as 
\begin{equation}
\Gamma ^{(0)}[\phi ]=S[\phi ],\text{ }\Gamma ^{(1)}[\phi ]=\frac{1}{2}Tr\ln
D^{-1},
\end{equation}
where 
\begin{equation}
D_{AB}^{-1}\equiv S_{AB}[\phi ]=\Delta _{AB}^{-1}+\frac{\delta
^{2}S^{int}[\phi ]}{\delta \phi _{A}\delta \phi _{B}}.
\end{equation}
The higher order effective action $\Gamma [\phi ]$ is given by the 1PI
vacuum diagrams with the propagator $D_{AB}^{-1}$ and the vertices $%
S_{A_{1\cdot \cdot \cdot }A_{N}}[\phi ]$ \cite{Jackiw}.

Now consider the functional identity

\begin{equation}
\frac{\delta J_{A}}{\delta \phi _{C}}\frac{\delta \phi _{C}}{\delta J_{B}}%
=\delta _{AB}.
\end{equation}
From (6), we can obtain

\begin{equation}
\frac{\delta J_{A}}{\delta \phi _{C}}=-\frac{\delta ^{2}\Gamma [\phi ]}{%
\delta \phi _{A}\delta \phi _{C}},
\end{equation}
and from (3),(5) and (7) we obtain

\begin{eqnarray}
\frac{\delta \phi _{C}}{\delta J_{B}} &=&\frac{\delta ^{2}W[J]}{\delta
J_{C}\delta J_{B}}=\frac{1}{\hbar }(\phi _{C}\phi _{B}-\frac{\int D\Phi \Phi
_{C}\Phi _{B}\ Exp[-\frac{1}{\hbar }(S(\Phi )-J_{A}\Phi _{A})]}{\int D\Phi \
Exp[-\frac{1}{\hbar }(S(\Phi )-J_{A}\Phi _{A})]})  \nonumber \\
&=&\frac{2}{\hbar }(\frac{\delta \Gamma ^{(0)}[\phi ]}{\delta \Delta
_{AC}^{-1}}-\frac{\delta \Gamma [\phi ]}{\delta \Delta _{AC}^{-1}}) 
\nonumber \\
&=&-\frac{2}{\hbar }\frac{\delta }{\delta D_{AC}^{-1}}\sum_{l=1}\hbar
^{l}\Gamma ^{(l)},
\end{eqnarray}
We have used the fact that $\Gamma ^{(n)}$ depends on $\Delta ^{-1}$ only
through the $D_{AB}^{-1}$ ( see (11) and (12) ) when $n\geq 1$ to obtain the
last line of the above equation. By using the identity

\begin{equation}
\frac{\delta }{\delta D_{AC}^{-1}}=\frac{\delta D_{PQ}}{\delta D_{AC}^{-1}}%
\frac{\delta }{\delta D_{PQ}}=-D_{AP}D_{CQ}\frac{\delta }{\delta D_{PQ}},
\end{equation}
and by substituting (14) and (15) into (13), we obtain

\begin{equation}
2\frac{\delta }{\delta D_{AB}}\sum_{l=1}\hbar ^{l}\Gamma ^{(l)}=-\hbar
D_{AM}^{-1}[\frac{\delta ^{2}\Gamma [\phi ]}{\delta \phi _{M}\delta \phi _{N}%
}]^{-1}D_{NB}^{-1}.
\end{equation}
By using (11), we can see that the order $\hbar $ term of (17) is already
satisfied. As usual, let us define the proper self-energy $\Pi $ and the
full propagator $G$ as

\begin{equation}
\Pi _{AC}\equiv \sum_{l=1}\hbar ^{l}\frac{\delta ^{2}\Gamma ^{(l)}[\phi ]}{%
\delta \phi _{A}\delta \phi _{C}}\equiv \sum_{l=1}\hbar ^{l}\Pi _{AC}^{(l)},
\end{equation}
and 
\begin{equation}
G_{AB}^{-1}\equiv \frac{\delta ^{2}\Gamma [\phi ]}{\delta \phi _{A}\delta
\phi _{B}}=D_{AB}^{-1}+\Pi _{AB}.
\end{equation}
so that 
\begin{equation}
G=D+D\sum_{l=1}(-\Pi D)^{l}.
\end{equation}
By substituting (18) and (20) into (17) and by multiplying $D_{AB}$, we
obtain the recursion relation for the effective action as 
\begin{equation}
2\frac{\delta \Gamma ^{(n)}}{\delta D_{AB}}D_{AB}=-Tr[\sum_{l=1}(-\Pi
D)^{l}]^{(n-1)}\text{ }(n\geq 2),
\end{equation}
where the notation [...]$^{(n)}$ means the order $\hbar ^{n}$ term of the
quantity inside of the bracket. Eq.(21) is the central result of this paper
and by using this equation, we can obtain the $n$-th order effective action
from the lower order self-energies. Note that the result of the operation $%
\frac{\delta \Gamma ^{(n)}}{\delta D_{AB}}D_{AB}$ is equal to multiplying
each diagrams in $\Gamma ^{(n)}$ by the number of the its propagators.

Now, let us apply the recursion relation (21) to the general bosonic field
theory which have vertices higher than the four-point vertex. In case of the
two-loop effective action $\Gamma ^{(2)}$, (21) becomes 
\begin{equation}
2\frac{\delta \Gamma ^{(2)}}{\delta D_{AB}}D_{AB}=\Pi _{AB}^{(1)}D_{BA}=%
\frac{\delta ^{2}\Gamma ^{(1)}[\phi ]}{\delta \phi _{A}\delta \phi _{B}}%
D_{BA}.
\end{equation}
The derivative with respect to $\phi $ can act either to the propagator $%
D_{AB}$ which contains the term $S_{AB}[\phi ]$ or to the vertex $%
S_{A_{1\cdot \cdot \cdot }A_{N}}[\phi ]$ as 
\begin{equation}
\frac{\delta D_{AB}}{\delta \phi _{C}}=-(D\frac{\delta D^{-1}}{\delta \phi
_{C}}D)_{AB}=-D_{AP}S_{CPQ}D_{QB},
\end{equation}
and 
\begin{equation}
\frac{\delta S_{A_{1\cdot \cdot \cdot }A_{N}}[\phi ]}{\delta \phi _{C}}%
=S_{A_{1\cdot \cdot \cdot }A_{N}C}[\phi ].
\end{equation}
In the graphical representation, a line represents the propagator $D$ and a $%
n-$point vertex have the factor $S_{A_{1\cdot \cdot \cdot }A_{n}}$. Also a
box with an capital letter represents the vertex which have indices that is
not contracted with the propagators attached to it so that 
\begin{equation}
\begin{picture}(130,40) \put(0,21){\framebox(20,10){}} \put(0,35){$A..B$}
\put(35,23){$=$}
\put(5,21){\line(-1,-1){10}}\put(10,21){\line(0,-1){13}}\put(20,21){%
\line(1,-1){10}} \put(15,15){$..$}
\put(-5,2){$P$}\put(12,2){$Q$}\put(28,2){$R$} \put(50,23){$S_{A..BP^{\prime
}Q^{\prime }..R^{\prime }}D_{PP^{\prime }} D_{QQ^{\prime}}..D_{RR^{\prime
}}$} \end{picture}.
\end{equation}
For example, (23) can be expressed as 
\begin{equation}
\frac{\delta }{\delta \phi _{C}}[\begin{picture}(30,20)
\put(5,3){\line(1,0){20}} \end{picture} ]_{AB}=-[\begin{picture}(60,20)
\put(5,3){\line(1,0){17}} \put(23,0){\framebox(6,6){}}\put(25,10){$C$}
\put(29,3){\line(1,0){20}} \end{picture} ]_{AB}.
\end{equation}
Then we can obtain from (11) and (18) 
\begin{eqnarray}
\Pi _{AB}^{(1)} &=&\frac{\delta ^{2}\Gamma ^{(1)}[\phi ]}{\delta \phi
_{A}\delta \phi _{B}}=\frac{1}{2}\frac{\delta }{\delta \phi _{A}}Tr[D\frac{%
\delta D^{-1}}{\delta \phi _{B}}]  \nonumber \\
&=&\frac{1}{2}[-D_{PQ}S_{AQR}D_{RS}S_{BSP}+D_{PQ}S_{ABPQ}]=\frac{1}{2}[-%
\begin{picture}(110,20) \put(15,0){\framebox(4,4){}}
\put(31,0){\framebox(4,4){}} \put(25,2){\circle{16}} \put(5,0){$A$}
\put(37,0){$B$} \put(53,0){$+$} \put(65,0){$AB$}
\put(85,0){\framebox(4,4){}} \put(95,2){\circle{16}} \end{picture} ],
\end{eqnarray}
By using (22), we can obtain 
\begin{equation}
\Gamma ^{(2)}[\phi ]=-\frac{1}{12}S_{AQR}S_{BSP}D_{AB}D_{PQ}D_{RS}+\frac{1}{8%
}S_{ABPQ}D_{AB}D_{PQ}=\begin{picture}(120,20) \put(5,0){$-\frac{1}{12}$}
\put(40,2){\circle{16}} \put(32,2){\line(1,0){16}}
\put(60,0){+$\frac{1}{8}$} \put(94,2){\circle{16}} \put(110,2){\circle{16}}
\end{picture}.
\end{equation}
In case of three-loop effective action $\Gamma ^{(3)}$, (21) becomes 
\begin{equation}
2\frac{\delta \Gamma ^{(3)}}{\delta D_{AB}}D_{AB}=(\Pi _{AB}^{(2)}-\Pi
_{AP}^{(1)}D_{PQ}\Pi _{QB}^{(1)})D_{BA}
\end{equation}
$\Pi ^{(2)}$ can be obtained from $\Gamma ^{(2)}[\phi ]$ by operating $\frac{%
\delta ^{2}}{\delta \phi _{A}\delta \phi _{B}}$. The graphical
representation of this operation to the diagrams of $\Gamma ^{(2)}[\phi ]$
is given by 
\begin{eqnarray}
\Pi _{AB}^{(2)} &=&\frac{\delta ^{2}}{\delta \phi _{A}\delta \phi _{B}}[-%
\frac{1}{12}\begin{picture}(20,20) \put(10,2){\circle{16}}
\put(2,2){\line(1,0){16}} \end{picture} +\frac{1}{8}\begin{picture}(40,20)
\put(10,2){\circle{16}} \put(26,2){\circle{16}} \end{picture} ]=\frac{\delta 
}{\delta \phi _{B}}[-\frac{1}{6}\begin{picture}(40,20) \put(5,0){$A$}
\put(15,0){\framebox(4,4){}} \put(25,2){\circle{16}}
\put(17,2){\line(1,0){16}} \end{picture} +\frac{1}{4}\begin{picture}(40,20)
\put(16,14){$A$} \put(16,8){\framebox(4,4){}} \put(20,2){\circle{16}}
\put(12,2){\line(1,0){16}} \end{picture} +\frac{1}{8}\begin{picture}(40,20)
\put(16,0){\framebox(4,4){}} \put(10,2){\circle{16}} \put(26,2){\circle{16}}
\put(16,10){$A$} \end{picture} -\frac{1}{4}\begin{picture}(50,20)
\put(10,0){\framebox(4,4){}} \put(2,0){$A$} \put(20,2){\circle{16}}
\put(36,2){\circle{16}} \end{picture} ]  \nonumber \\
&=&-\frac{1}{6}\begin{picture}(45,20) \put(8,0){$AB$}
\put(25,0){\framebox(4,4){}} \put(35,2){\circle{16}}
\put(27,2){\line(1,0){16}} \end{picture} -\frac{1}{6}\begin{picture}(50,20)
\put(5,0){$A$} \put(38,0){$B$} \put(15,0){\framebox(4,4){}}
\put(31,0){\framebox(4,4){}} \put(25,2){\circle{16}}
\put(17,2){\line(1,0){16}} \end{picture} +\frac{1}{2}(\begin{picture}(40,20)
\put(2,0){$B$} \put(20,14){$A$} \put(20,8){\framebox(4,4){}}
\put(14,0){\framebox(4,4){}} \put(24,2){\circle{16}}
\put(16,2){\line(1,0){16}} \end{picture} +\begin{picture}(40,20)
\put(2,0){$A$} \put(20,14){$B$} \put(20,8){\framebox(4,4){}}
\put(14,0){\framebox(4,4){}} \put(24,2){\circle{16}}
\put(16,2){\line(1,0){16}} \end{picture})+\frac{1}{4}\begin{picture}(30,20)
\put(7,14){$AB$} \put(11,8){\framebox(4,4){}} \put(15,2){\circle{16}}
\put(7,2){\line(1,0){16}} \end{picture} -\frac{1}{2}\begin{picture}(30,30)
\put(13,-5){\framebox(4,4){}} \put(12,-14){$B$} \put(15,5){\circle{16}}
\put(13,11){\framebox(4,4){}} \put(12,17){$A$} \put(7,5){\line(1,0){16}}
\end{picture} -\frac{1}{2}\begin{picture}(30,20) \put(7,14){$A$}
\put(20,14){$B$} \put(9,6){\framebox(4,4){}} \put(19,6){\framebox(4,4){}}
\put(15,2){\circle{16}} \put(7,2){\line(1,0){16}} \end{picture}  \nonumber \\
&&+\frac{1}{8}\begin{picture}(40,20)
\put(16,0){\framebox(4,4){}}\put(10,2){\circle{16}} \put(26,2){\circle{16}}
\put(12,10){$AB$} \end{picture} -\frac{1}{4}(\begin{picture}(50,20)
\put(10,0){\framebox(4,4){}} \put(2,0){$B$} \put(26,10){$A$}
\put(26,0){\framebox(4,4){}} \put(20,2){\circle{16}} \put(36,2){\circle{16}}
\end{picture} +\begin{picture}(50,20) \put(10,0){\framebox(4,4){}}
\put(2,0){$A$} \put(26,10){$B$} \put(26,0){\framebox(4,4){}}
\put(20,2){\circle{16}} \put(36,2){\circle{16}} \end{picture})-\frac{1}{4}%
\begin{picture}(55,20) \put(42,0){$AB$} \put(36,0){\framebox(4,4){}}
\put(14,2){\circle{16}} \put(30,2){\circle{16}} \end{picture} +\frac{1}{4}%
\begin{picture}(60,20) \put(10,0){\framebox(4,4){}} \put(2,0){$A$}
\put(20,2){\circle{16}} \put(42,0){\framebox(4,4){}} \put(48,0){$B$}
\put(36,2){\circle{16}} \end{picture} +\frac{1}{2}\begin{picture}(45,30)
\put(18,-5){\framebox(4,4){}} \put(17,-14){$B$} \put(20,5){\circle{16}}
\put(18,11){\framebox(4,4){}} \put(17,17){$A$} \put(36,5){\circle{16}}
\end{picture}.
\end{eqnarray}
By substituting $\Pi ^{(1)}$ and $\Pi ^{(2)}$ given in (27) and (30) into
(29), we obtain 
\begin{eqnarray}
\Gamma ^{(3)}[\phi ] &=&-\frac{1}{16}\begin{picture}(50,20)
\put(10,2){\circle{16}} \put(26,2){\circle{16}} \put(42,2){\circle{16}}
\end{picture} -\frac{1}{48}\begin{picture}(40,20) \put(15,3){\circle{20}}
\put(5,3){\line(2,1){10}} \put(5,3){\line(2,-1){10}}
\put(25,3){\line(-2,1){10}} \put(25,3){\line(-2,-1){10}} \end{picture} +%
\frac{1}{8}\begin{picture}(40,20) \put(15,3){\circle{20}}
\put(15,-7){\line(1,2){8}}\put(15,-7){\line(-1,2){8}} \end{picture} +\frac{1%
}{8}\begin{picture}(40,20) \put(15,3){\circle{20}} \put(35,3){\circle{20}}
\put(15,13){\line(0,-1){20}}\end{picture}  \nonumber \\
&&-\frac{1}{16}\begin{picture}(40,20) \put(15,3){\circle{20}}
\put(10,12){\line(0,-1){18}} \put(20,12){\line(0,-1){18}} \end{picture} -%
\frac{1}{24}\begin{picture}(40,20) \put(15,3){\circle{20}}
\put(15,13){\line(0,-1){10}} \put(5,3){\line(1,0){20}} \end{picture} -\frac{1%
}{12}\begin{picture}(60,20) \put(15,3){\circle{20}} \put(35,3){\circle{20}}
\put(5,3){\line(1,0){20}}\end{picture} +\frac{1}{48}\begin{picture}(40,20)
\put(12,0){\circle{20}} \put(32,0){\circle{20}} \put(22,3){\line(1,2){8}}
\put(22,3){\line(-1,2){8}} \put(14,20){\line(1,0){16}} \end{picture}
\end{eqnarray}
.

In the case of the four-loop effective action $\Gamma ^{(4)}$, (21) becomes 
\begin{equation}
2\frac{\delta \Gamma ^{(4)}}{\delta D_{AB}}D_{AB}=(\Pi _{AB}^{(3)}-2\text{ }%
\Pi _{AP}^{(2)}D_{PQ}\Pi _{QB}^{(1)}+\Pi _{AP}^{(1)}D_{PQ}\Pi
_{QR}^{(1)}D_{RS}\Pi _{SB}^{(1)})D_{BA},
\end{equation}
It is straightforward to obtain $\Gamma ^{(4)}$ by following the same steps
as before. The result is 
\begin{equation}
\Gamma ^{(4)}=\Gamma _{1PI}^{(4A)}+\Gamma _{2PI}^{(4B)}+\Delta \Gamma ^{(4)},
\end{equation}
where $\Gamma _{1PI}^{(4A)}$ is the Feynman diagrams of the four-loop 1PI
effective action obtained from the three and four-point vertex and $\Gamma
_{2PI}^{(4B)}[\phi ]$ is the Feynman diagrams of the four-loop 2PI effective
action obtained from the higher vertices. $\Gamma _{1PI}^{(4A)}[\phi ]$ and $%
\Gamma _{2PI}^{(4B)}[\phi ]$ have been reported previously \cite{R7,2PI} and
we have obtained a result which agree with the previous results exactly. $%
\Delta \Gamma ^{(4)}$ which is the Feynman diagrams of the four-loop 1PI
effective action obtained from the higher vertices were not reported
previously and is given by

\begin{eqnarray}
\Delta \Gamma ^{(4)} &=&-\frac{1}{24}\begin{picture}(50,20)
\put(15,3){\circle{20}} \put(35,3){\circle{20}}
\put(15,13){\line(0,-1){20}}\put(25,3){\line(1,0){20}} \end{picture} +\frac{1%
}{24}\begin{picture}(65,20) \put(12,3){\circle{20}} \put(32,3){\circle{20}}
\put(52,3){\circle{20}} \put(2,3){\line(1,0){20}} \end{picture} -\frac{1}{8}%
\begin{picture}(55,20) \put(12,3){\circle{20}} \put(32,3){\circle{20}}
\put(5,10){\line(1,0){15}} \put(2,3){\line(1,0){20}} \end{picture} +\frac{1}{%
12}\begin{picture}(55,25) \put(12,3){\circle{20}} \put(32,3){\circle{20}}
\put(12,18){\circle{10}} \put(2,3){\line(1,0){20}} \end{picture}  \nonumber
\\
&&+\frac{1}{24}\begin{picture}(65,20) \put(12,3){\circle{20}}
\put(32,3){\circle{20}} \put(52,3){\circle{20}} \put(22,3){\line(2,1){15}}
\end{picture} +\frac{1}{32} \begin{picture}(40,20) \put(12,0){\circle{20}}
\put(32,0){\circle{20}} \put(22,3){\line(1,2){8}} \put(22,3){\line(-1,2){8}}
\put(14,20){\line(1,0){16}} \put(12,10){\line(0,-1){20}} \end{picture} -%
\frac{1}{32} \begin{picture}(75,20) \put(12,0){\circle{20}}
\put(32,0){\circle{20}} \put(22,3){\line(1,2){8}} \put(22,3){\line(-1,2){8}}
\put(14,20){\line(1,0){16}} \put(52,0){\circle{20}} \end{picture}
\end{eqnarray}

\section{ Discussions and Conclusions}

In this paper, we have derived a new kind of recursion relation to obtain
the effective action. We have applied this method to the general bosonic
field theory which have vertices higher than the four-point vertex and have
obtained the graphical representation of the four-loop effective action. The
1PI diagrams of the $\phi ^{4}$-theory with only three and four-point
vertices agreed with previous results and we have given the results for the
1PI diagrams with the higher vertices. The extension of the method we have
used in this paper to obtain the recursive generation of the 1PI effective
action to the case of the 2PI and 4PI effective action is in progress.

\begin{acknowledgments}
This research was supported in part by the Institute of Natural Science.
\end{acknowledgments}
\bibliographystyle{plain}
\bibliography{ee}

\begin{thebibliography}{99}
\bibitem{Sher}  For a review and references, see M. Sher, Phys. Rep. 179,
273 (1989).

\bibitem{Jackiw}  R. Jackiw, Phys. Rev. D9, 1686 (1974) .

\bibitem{F}  J.Kubldeck, M.Bohm and A.Denner, Comput. Phys. Commum. 60, 165
(1990) ; T. Hahn, hep-ph/0012260, http://www.feynarts.de/.

\bibitem{Q}  P.Nogueira, J. Comput. Phys. 105, 279 (1993); ftp://gtae2.
ist.utl.pt/pub/qgraf/.

\bibitem{R2}  H. Kleinert, A. Pelster, Phys.Rev. D61, 085017 (2000).

\bibitem{R3}  B. Kastening, Phys. Rev. E 61, 3501 (2000).

\bibitem{R4}  H. Kleinert, A. Pelster, B. Kastening, M. Bachmann , Phys.Rev.
E62, 1537 (2000).

\bibitem{R5}  A. Pelster, H. Kleinert, M. Bachmann , Annals Phys. 297, 363
(2002).

\bibitem{R6}  H. Kleinert, A. Pelster, B. Van den Bossche ,Physica A312, 141
(2002).

\bibitem{R7}  A. Pelster, H. Kleinert, Physica A323, 370 (2003)

\bibitem{R8}  A. Pelster, K. Glaum, Physica A335, 455 (2004).

\bibitem{2PI}  K. Kajantie, M. Laine and Y. Schroder, Phys.Rev. D65 045008
(2002).

\bibitem{4PI}  C.K.Kim, Phys.Rev.D72, 085007 (2005); J. Berges,
Phys.Rev.D70, 105010 (2004) and M.E.Carrington, Eur. Phys.J.C35, 383 (2004).

\bibitem{lattice}  K. Farakos, K.Kajantie, Rummukainen and Y.Schroder, Nucl.
Phys. B442, 317 (1995).

\bibitem{Coleman}  S. Coleman and E. Weinberg, Phys. Rev. D7, 1888 (1973).
\end{thebibliography}

\end{document}